\documentclass{an}
\usepackage{graphicx}
\usepackage{times}
\Volume{00}              
\Year{0000}              
\Month{00}               
\Pagespan{000}{000}      

\begin{document}
\lhead[\thepage]{A.N. S.F.S\'anchez: The merging/AGN connection}
\rhead[Astron. Nachr./AN~{\bf XXX} (200X) X]{\thepage}
\headnote{Astron. Nachr./AN {\bf 32X} (200X) X, XXX--XXX}

\title{The merging/AGN connection: A case for 3D spectroscopy}

\author{S.F. S\'anchez\inst{1} \and L. Christensen\inst{1} \and
  T. Becker\inst{1} \and A. Kelz\inst{1} \and K. Jahnke\inst{1} \and
  C.R. Benn\inst{2} \and B. Garc\'\i a-Lorenzo\inst{3} \and M.M. Roth\inst{1}}
\institute{Astrophysikalisches Institut Potsdam, And der Sternwarte 16, 14482
  Potsdam, Germany \and Isaac Newton Group, Apt. 321, S/C de La Palma, Spain
\and  Instituto de Astrof\'\i sica de Canarias, 3805 La Laguna, Tenerife, Spain}
\date{Received {date will be inserted by the editor}; 
accepted {date will be inserted by the editor}} 

\abstract{We discuss an ongoing study of the connection between galaxy
merging/interaction and AGN activity, based on integral field spectroscopy.
We focus on the search for AGN ionization in the central regions of 
mergers, previously not classified as AGNs.
We present here the science case, the current status of the project, and plans
for future observations. 
}
\correspondence{ssanchez@aip.de}

\maketitle

\section{Introduction}

It is known that interactions between galaxies can deeply affect their
evolution and stellar populations. Interacting galaxies are more active in the
UV (\cite{lt78}), in the near-infrared (\cite{jw85}), in optical emission-line
strength (\cite{kk84}), and in radio emission (\cite{sto78}; \cite{hu81}). It
is also known that ultraluminous infrared galaxies (ULIRGs), a class of
objects with luminosities above 10$^{\rm 12}$L$_{\odot}$ in the Far Infrared
(FIR), are interacting or merging systems (\cite{sd88}; \cite{cle96}).

There is also growing evidence that merging processes could be the
triggering/fuelling mechanism of AGN activity (\cite{cs01}; \cite{jan02};
\cite{sg03}). Some studies found that Seyfert galaxies are often interacting
systems (e.g. \cite{ke85}), although this is not always the case (\cite{da85};
\cite{bu86}). Quasars often inhabit rich galaxy environments and/or have
nearby companions (e.g. \cite{sg02}) and quasar hosts exhibit signs of
encounters and mergers in the form of tidal debris and multiple nuclei
(\cite{cs01}; \cite{sg03}). Merging and interactions appear to be an efficient
process for transferring material to the inner kpc or even pc of a galaxy
(e.g.  \cite{mh96}).

An evolutionary scenario has been suggested, in which major mergers between
galaxies trigger intense star formation processes, and may allow the infall of
gas into the nuclear regions. Both the AGN and the starburst heat the dust,
and the object is observed as a ULIRG. It is known that ULIRGs have the same
bolometric luminosities and space densities as QSOs, which supports the
hypothesis of being their progenitors (\cite{sd88}). The presence of both
starburst and AGN activity in ULIRGs naturally raises the question of their
evolutionary connection. It is expected that a family of transition objects
between ULIRGs and AGNs exists. These transition objects would be {\it either}
AGNs with host galaxies showin strong evidence of recent interaction/merging
{\it or} merging systems, luminous in the FIR, with strong star formation
which harbor a faint AGN in the inner regions. This AGN could be obscured
(\cite{sd88}; \cite{ko03}), or blurred by the surrounding star forming region
(\cite{dp00}; \cite{lu99}).

For the most violent cases of AGN activity, i.e., the QSOs, the results are
not conclusive. Many authors claim to find possible interaction/merger traces
in QSO hosts (e.g., \cite{hu94}; \cite{ba97}; \cite{sg03}). However, Dunlop et
al. (2003) show no evidence for a connection between ULIRGs and their sample
of low-$z$ AGN hosts. Their hosts present a pure elliptical surface brightness
profile, with no evidence of interactions, and they are located in a different
region of the fundamental plane from the ULIRGs. This makes improbable an
evolution from the latter to the former population based on merging processes.
On the other hand, Canalizo \& Stockton (2001) found via an imaging and
spectroscopic analysis that all {\it transition} QSOs are undergoing tidal
interactions or are major mergers.  They selected their sample of QSOs to have
similar FIR properties to the ULIRGs (see below).  They conclude that at
least a family of AGNs evolves from a ULIRG progenitor, formed by a merger of
galaxies.  It is still an open question whether the host galaxies of these
transition QSOs evolve later to galaxies without trace of recent
interactions, or if the two families are different.

\begin{figure}
\resizebox{\hsize}{!}
{\includegraphics[width=7cm,angle=-90]{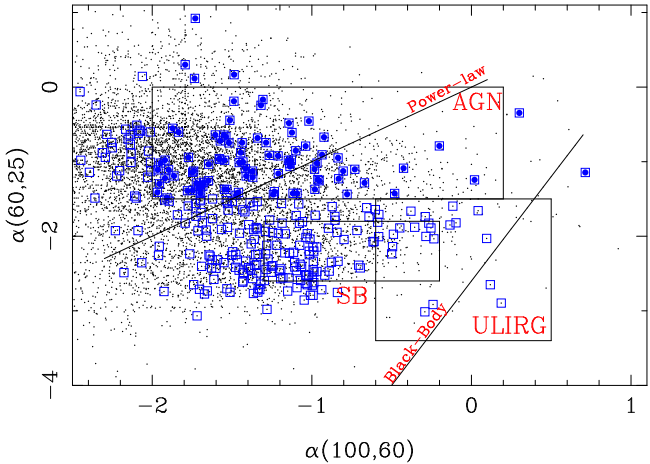}}
\caption{Far infrared spectral indeces between 25$\mu$m and 60$\mu$m versus
  spectral indexes between 60$\mu$m and 100$\mu$m for the IRAS sources with
  known morphology (dots). The open squares shows the 312 objects with known
  redshift, and the solid squares shows our sample of 88 objects.
  The boxes show the location of AGNs, ULIRGs and star-forming galaxies in this 
  plane. We also show the location in the plane of spectral  indices
  corresponding to black bodies and power laws (solid lines). }
\label{fig1}
\end{figure}

\section{The Project}

We have started a project with the aim of looking for {\it transition} objects
between merger galaxies and AGNs. We selected a sample of merging galaxies
from the IRAS catalogue of 472 sources with known morphological properties
(\cite{iras88}; \cite{iras89}). Of these, we have selected the 312 sources
with published redshifts. This sample contains galaxies whose morphology shows
evidences of recent merger events, i.e., tails, plumes, rings or other
morphological features consistent with strong tidal interaction.  These
systems are being observed at late phases of the interaction, since severe
morphological disturbances, such as long tidal tails, require $>$10$^8$ yr to
develop (\cite{tt72}; \cite{mh96}).

\begin{figure}
\resizebox{\hsize}{!}
{\includegraphics[width=7cm]{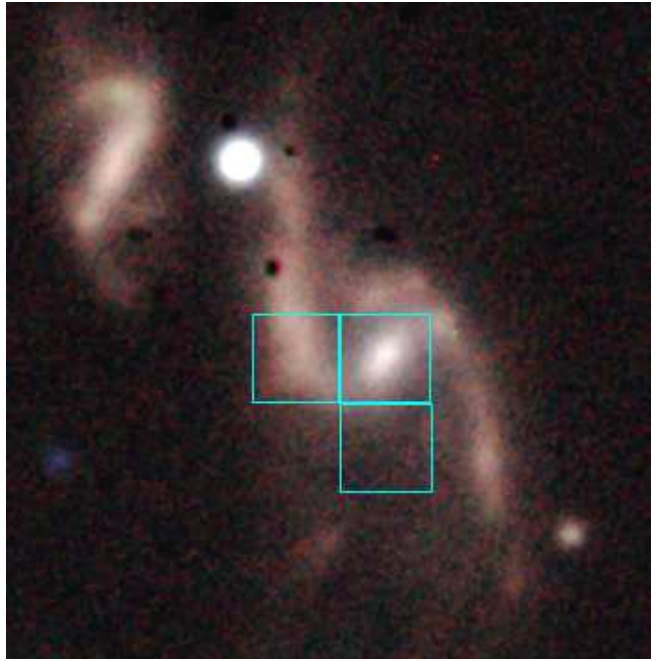}} 
\caption{True-color image of the object, IRAS 12110+1624, created using  $B$, $V$, and $R$ band images obtained with the PMAS AG camera. The field of view is $\sim$90$\arcsec$$\times$90$\arcsec$}
\label{point}
\end{figure}

Figure \ref{fig1} shows the distribution of far infrared (FIR) spectral
indices for the IRAS catalogue ($\alpha_{60}^{100}$ vs. $\alpha_{25}^{60}$),
together with our selected sample (see below). We overplotted the location of
the AGNs, ULIRGs and galaxies with large starformation rate (SFR) and the
location of two typical spectral energy distributions of the FIR emission,
power-law, typical for AGNs, and black-body thermal emission, typical for
ULIRGs and galaxies with large SFR. The FIR emission is produced by dust that
absorbs optical-UV flux produced by recently formed stars and/or AGNs. The
close association of these stars (or AGNs) with the gas and dust from which
they formed (or feed), assures that much of the luminosity of the group of
young stars (or AGN) will emerge in the infrared.  Therefore, the FIR emission
traces the most violent processes of activity in galaxies (\cite{so84};
\cite{kk87}; \cite{sd88}; \cite{bu88}; \cite{sm96}).

\begin{table*}
\caption{Summary of the properties of the subsample of observed objects}
\label{tab1}
\begin{tabular}{crrcrrrrl}
\hline
IRAS Name  & $z$ & $B$-mag& Class. & f$_{12}$ & f$_{25}$ & f$_{60}$ & f$_{100}$ & Instrument\\
\hline
10219-2828  & 0.02677 &15.1 & Double  &   0.25 &   0.93 &    4.82 &    5.78 & VIMOS LR Red \\
10484-0153  & 0.01450 &12.9 & Peculiar  &   0.34 &   0.57 &    4.74 &    9.67 & VIMOS LR Blue/Red\\
11096-4738  & 0.01438 &     & Double? &   0.36 &   0.25 &    0.66 &    1.80 & VIMOS LR Blue/Red\\
11500-0211  & 0.00345 &14.1 & Peculiar  &   0.28 &   0.34 &    1.01 &    1.00 & VIMOS LR Blue/Red\\
12042-3140  & 0.02309 &     & Double  &   0.29 &   0.73 &    7.30 &   12.39 & VIMOS LR Blue/Red\\
12110-3412  & 0.00924 &14.2 & Double  &   0.35 &   0.38 &    1.71 &    2.82 & VIMOS LR Blue/Red\\
12110+1624  & 0.02379 &14.8 & Mult. system &   0.40 &   0.27 &    0.52 &    1.13 & PMAS Mosaic V300\\
12193-3942  & 0.07454 &     & Double  &   0.53 &   0.25 &    0.83 &    1.37 & VIMOS LR Blue/Red\\
12447-5316  & 0.00606 &     & Double? &   0.26 &   0.26 &    2.37 &    3.99 & VIMOS LR Blue/Red\\
12583-3619  & 0.01622 &15.5 & Double? &   0.25 &   0.28 &    0.62 &    1.00 & VIMOS LR Blue/Red\\
13031-5717  & 0.01961 &     & Double  &   0.25 &   0.42 &    2.92 &    5.16 & VIMOS LR Blue/Red\\
13473-4801  & 0.01075 &     & Double? &   0.25 &   0.95 &    5.64 &    6.36 & VIMOS LR Blue/Red\\
13477-4848  & 0.01010 &12.6 & Double? &   0.67 &   1.67 &   13.09 &   23.05 & VIMOS LR Blue/Red\\
14442-1902  & 0.07403 &     & Double  &   0.25 &   0.40 &    0.97 &    2.04 & VIMOS LR Blue/Red\\
15305-0127  & 0.00927 &14.5 & S-Irregular   &   0.25 &   0.27 &    0.84 &    2.28 & VIMOS LR Blue/Red\\
16109+6042  & 0.01376 &14.5 & Double system &   0.25 &   0.36 &    0.38 &    1.33 & PMAS Mosaic V300\\
16176-6325  & 0.01721 &     & Double  &   0.35 &   0.79 &    3.86 &    5.21 & VIMOS LR Blue/Red\\
16229-6640  & 0.02179 &     & Triple  &   0.27 &   0.50 &    4.73 &    9.53 & VIMOS LR Blue/Red\\
16330-6820  & 0.04697 &     & Double  &   0.19 &   0.74 &    7.16 &   12.37 & VIMOS LR Blue/Red\\
16365+4202  & 0.02707 &15.5 & Double system &   0.25 &   0.25 &    0.51 &    1.61 & PMAS Mosaic V300\\
\hline
\end{tabular}

Columns list: (1) the IRAS name of the object, (2) the redshift, (3) the
$B$-band magnitude, (4) the classification in the IRAS catalogue, (5-8) the
IRAS flux at 12$\mu$m, 25$\mu$m, 60$\mu$m and 100$\mu$m and (9) the
instrument/setup with which it was observed.

\end{table*}

The degree of activity (SFR or AGN activity) increases with
$\alpha_{60}^{100}$ and the thermal emission (\cite{bu88}). Therefore, we
selected the 88 objects with warm emission in the far infrared in a similar
way to Canalizo \& Stockton (2001), but focused on the interacting systems,
i.e.  excluding all the objects previously classified as AGNs. The final
sample comprises the interacting/merging objects with $\alpha_{60}^{100}>-$2
and $\alpha_{25}^{60}>-$1.5 \footnote{Spectral index defined as
  $F_\nu=\nu^{-\alpha}$} (solid squares on Fig. \ref{fig1}). We have started
to study these objects using Integral Field Spectroscopy (IFS), looking for
AGN activity and trying to understand the connection of this activity and
galaxy merging/interaction.

There have been several attempts to disentangle the relation between galaxy
interactions and galaxy activity (if any).  Many of these studies were focused
on the understanding of the triggering of starformation by tidal interaction
(\cite{lt78}; \cite{kk87}; \cite{bu88}; \cite{sm96}), and most of them were
based on optical/NIR imaging and slit-spectroscopy (\cite{kk87}; \cite{bu88};
\cite{bu02}; \cite{dp00}). They all found an increase of the SFR for
merger/interacting galaxies, especially in the nuclear regions (see
\cite{ber03} for a counter-example). The specific problem of the
AGN/interaction relation has been partially addressed by these studies.
Contrary to expectations, they found a deficiency of AGNs among interacting or
merging systems compared to isolated galaxies (\cite{da85};
\cite{ke85};\cite{bu88}; \cite{dp00}; \cite{sw92}; \cite{lk95}). The fraction
of AGNs increases with FIR luminosity, being about $\sim$30\% for the extreme
case of the ULIRGs (\cite{bu02}).

The lack of AGN-like emission detected could be due to the fact that some of
the galaxies have strong circumnuclear star-forming regions (\cite{ke85};
\cite{kk87}; \cite{dp00}). Therefore, the H{\rm II}-region spectra would
dilute the AGN spectra when the observations are made at a low spatial
resolution, as is the case in nearly all the previous studies. Even in the
case of good spatial resolution, the uncertainty of the location of the
nucleus, makes slit spectroscopy a dangerous method to search for faint AGN
activity possibly diluted by nearby star formation. This problem can easily be
addressed by IFS. Our study will yield spatially resolved spectroscopic
information of the objects, together with better spatial resolution than
previous studies.

The number of IFS studies over representative samples of merger galaxies is
small (e.g. \cite{ch98}). To our knowledge, only one large systematic program
is currently ongoing, focused on the study of the evolution of ULIRGs from
cool and warm infrared emitters (\cite{ac03}, and references therein), and the
overall properties of these objects. The different selection criteria for our
sample, and the different focus, make the two studies complementary for the
overall undertanding of the connection between interactions and AGN activity.

\begin{figure*}
\resizebox{\hsize}{!}
{\includegraphics[width=1.0\textwidth,viewport=70 570 610 800,clip]{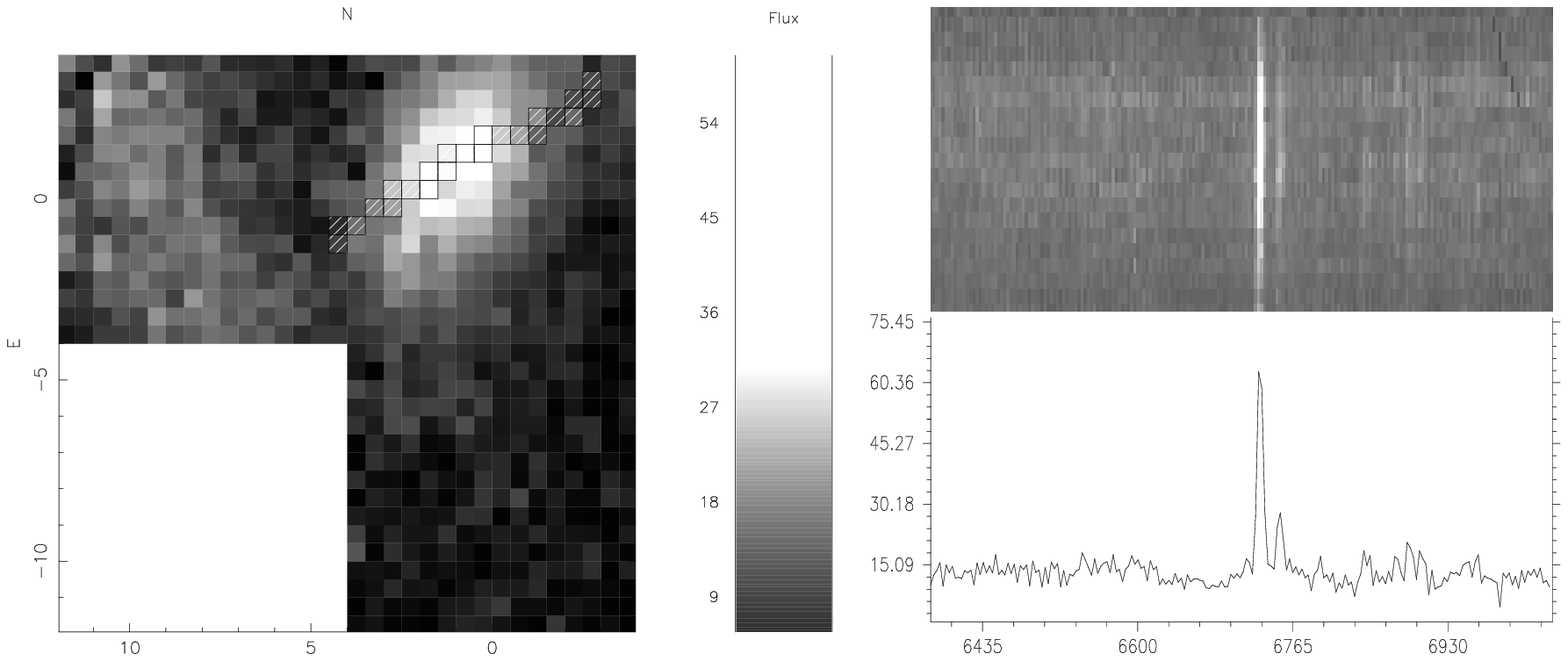}}
\caption{{\bf Left:} Polychromatic of the IRAS 12110+1624 data taken with
  PMAS, averaged between
  6689\AA\ and 6771\AA, including the H${\alpha}$ emission line (at
  $\sim$6728\AA)
The dashed squares show the selected spaxels for the construction of a pseudo
slit-spectrum. {\bf Right:} Section of the pseudo slit-spectrum, centered on
the H$\alpha$ region, created by the previous selection of spaxels. The top
panel shows the row-stacked spectra, spaxel-to-spaxel, from NW to SE (see left
panel). The bottom panel shows the averaged spectrum, flux in counts and
wavelength range in \AA.}
\label{example}
\end{figure*}

\section{Current State of the Project}

We have started an observational program using different telescopes and
instruments, to obtain low-resolution IFS of a representative subsample of our
selection. The wavelength ranges of interest are around H${\alpha}$ and
H${\beta}$, in order to obtain line ratios (e.g. [O${\rm III}$]/H${\beta}$ and
[N${\rm II}$]/H${\alpha}$) which allow us to determine the origin of the
ionization (\cite{vo87}). Dust extinction will be estimated from the
H${\alpha}$/H${\beta}$ ratio. Fitting models to the strongest emission lines,
spaxel to spaxel, would allow us to determine the kinematic 2D structure with
an accuracy of $\sim$20 km s$^{-1}$ (for low-resolution spectroscopy,
$\sim$6\AA). This accuracy is good enough to detect gas inflows/outflows
induced by the interaction (on the order of $\sim$100-200 km s$^{-1}$), which
could trigger starformation/AGN activity in the nuclear regions (e.g.
\cite{ac02}).

The average projected size of our objects is $\sim$40$\arcsec$, larger than
the field-of-view of actual Integral Field Units (IFUs), apart from VIMOS.
This forces us to do mosaic pointings around the central areas. Figure
\ref{point} shows an example of this mosaicing technique. This true-color
image of IRAS 12110+1624, was created using $B$, $V$ and $R$ band images
obtained with the PMAS A\&G camera.  This camera is a unique feature of PMAS,
that allows one to obtain $\sim$3$\arcmin$$\times$3$\arcmin$ field of view CCD
images centred on the target (\cite{ro00}). The three squares show the
position of the PMAS mosaic pointings for the object. Figure \ref{example}
shows the same regions observed with the PMAS spectrograph as a polychromatic
cut of the H$\alpha$ region together with a pseudo slit-spectrum created by
selecting certain spaxels across the central region of the object\footnote{We
  define spaxel as an spatial element of an IFU: lens, fiber, ...}. The data
were reduced using P3d (\cite{be01}). This figure, created using the Euro3D
visualization tool (\cite{san03}), shows the unique possibilities of IFS for
the analysis of these objects. A complete analysis of these data will be
presented elsewhere (\cite{san03_2}).

Table \ref{tab1} summarizes the current state of the project. To date we
have obtained IFS of 20 objects of the sample, 17 of them with VIMOS at the
VLT, and 3 more observed using PMAS at the 3.5m CAHA telescope. We have 3
nights allocated on the WHT next semester for further
observations using INTEGRAL (\cite{ar98}).

\acknowledgements

This project is founded by the Euro3D Training Network on Integral Field
Spectroscopy, funded by the European Commission under contract No.
HPRN-CT-2002-00305. L.Christensen acknowledges support by the German
Verbundforshung associated with the ULTROS project, grant no. 05AE2BAA/4.


\end{document}